\documentclass[twocolumn]{aastex61}

\newcommand\aastex{AAS\TeX}

\received{}
\revised{}
\accepted{}

\shorttitle{\aastex\ Extra-tidal stars around M53 and NGC 5053}
\shortauthors{Chun et al.}

\begin{document}

\title{Extra-tidal stars and chemical abundance properties of two metal-poor globular clusters M53 (NGC 5024) and NGC 5053}

\correspondingauthor{Sang-Hyun Chun}
\email{shyunc@kasi.re.kr}
\author[0000-0002-6154-7558]{Sang-Hyun Chun}
\affiliation{Korea Astronomy and Space Science Institute, 776 Daedeokdae-ro, Yuseong-gu, Daejeon 34055, Republic of Korea} 
\author{Jae-Joon Lee}
\affiliation{Korea Astronomy and Space Science Institute, 776 Daedeokdae-ro, Yuseong-gu, Daejeon 34055, Republic of Korea}
\author{Dongwook Lim}
\affiliation{Zentrum f\"ur Astronomie der Universit\"at Heidelberg, Astronomisches Rechen-Institut, M\"onchhofstr. 12-14, 69120 Heidelberg, Germany}




\begin{abstract}
We search for extra-tidal stars around two metal-poor Galactic globular clusters, M53 and NGC 5053, using the near-infrared APOGEE spectra. Applying the t-SNE algorithm on the chemical abundances and radial velocities results in identification of two isolated stellar groups composed of cluster member stars in the t-SNE projection plane. With additional selection criteria of radial velocity, location in the color-magnitude diagram, and abundances from a manual chemical analysis, we find a total of 73 cluster member candidates; seven extra-tidal stars are found beyond the tidal radii of the two clusters. The extra-tidal stars around the clusters tend to be located along the leading direction of the cluster proper motion, and the individual proper motion of these stars also seems to be compatible to those of clusters. Interestingly, we find that one extra-tidal star of NGC 5053 is located on the southern outskirts of M53, which is part of common stellar envelope by the tidal interaction between two clusters. 
We discuss the nature of this star in the context of the tidal interaction between two clusters. We find apparent Mg-Al anticorrelations with a clear gap and spread ($\sim$0.9 dex) in Al abundances for both clusters, and a light Si abundance spread ($\sim$0.3 dex) for NGC 5053. 
Since all extra-tidal stars have Mg enhanced and Al depleted features, they could be first-generation stars of two globular clusters. 
Our results support that M53 and NGC 5053 originated in dwarf galaxies and are surrounded by extended stellar substructures of more numerous populations of clusters.

\end{abstract}




\section{Introduction}\label{sec:intro}
Merging and accretion events of small fragments, such as dwarf satellite galaxies into the Milky Way, necessarily leave tidal tails and
stellar streams, which help us to understand the dynamical evolution and formation history of the Galaxy
~\citep[e.g.,][]{Ibata1994,Belo2006,Koch2012,Helmi2018,shipp2018,Massari2019}. Such streams are also known to be
associated with globular clusters. Some globular clusters in the Milky Way
are considered to be the first building blocks of the Galaxy. Furthermore, all globular clusters
indeed lose their mass through tidal disruption and dynamical friction~\citep{Fall1977,Fall1985,Gnedin1997,Baum2003}.
Some globular clusters show clear tidal tails or extended sub-halos in their vicinity~\citep{Oden2001,Jordi2010,Myeong2017,Kuzma2018},
which implies that part of the stars that consist of the Milky Way halo~\citep[from $11 \%$ to $50 \%$ depending on the assumptions;][]{Mackey2004, Martell2010, Koch2019} came from globular clusters.
A lot of effort has been and is still being made to find such stars that originate from globular clusters~\citep[e.g.,][]{Anguiano2016,Fern2016a,Navin2016,Minniti2018,Kundu2019}.
Such studies are very important to understand the formation and evolution
of the Milky Way; thus, finding more stars that originated from globular clusters in the halo field is necessary.

Multiple population and light element anomalies of globular clusters are useful signatures to 
identify globular cluster-origin (GC-origin) stars in the Milky Way. For example, they show distinctive chemical patterns like C-N, O-Na, and 
Mg-Al anticorrelation that
are unique among globular clusters~\citep[e.g.,][]{Gratton2004,Sneden2004,Carretta2009,Meszaros2015}. 
Such chemical anomalies that are observed among globular clusters can be used to distinguish GC-origin stars from normal field halo stars.
Sky survey projects, such as Sloan digital Sky Survey (SDSS), have found several field giants with
atypical chemical patterns similar to those of second-generation populations in globular clusters
~\citep[e.g.,][]{Gilmore2012,Fern2016b,Martell2016,Majew2017,Schi2017b}, and the contribution of globular clusters to
the formation of the Galactic halo and bulge is being actively discussed.

In this work, we focus on stars in the vicinity of two globular clusters in the Galactic halo, M53 (NGC 5024) and NGC 5053,
and we search for extra-tidal stars of the two clusters.
These two clusters are among the most metal-poor clusters in the Milky Way~\citep[$\mathrm{[Fe/H]=-2.10}$ for M53 and $\mathrm{[Fe/H]=-2.27}$ for NGC 5053;][adopted from~\citealp{Harris1996}]{Searle1978,Sunt1988,Geisler1995,Carretta2009}.
They are located within $1\degr$ on the projected sky, and the distance between
the clusters is only $\sim500$ pc. 
{Due to their proximity in the sky, the physical association between the two clusters and their origin
have been a subject of study. 
~\citet{Forbes2010} discussed the possibility that one or both of them are the nucleus of a disrupted dwarf galaxy. 
Since they are along the Sagittarius (Sgr) streams, their possible association with the Sgr Dwarf Spheroidal Galaxy (dSph) have
long been suspected~\citep{Palma2002,Bella2003,Law2010}.
Recent studies of accurate proper motions and orbit calculations have shown that
the orbits of these two clusters are significantly different from that of Sgr dSph, and thus 
excluded their association~\citep{Sohn2018, Tang2018}.
On the other hand,  ~\citet{Yoon2002} found that the seven globular clusters with the lowest metallicity ($\mathrm{[Fe/H]}<-2.0$),
including M53 and NGC5053, display a spatial alignment of which plane is perpendicular to the line joining the present position of the Sun and the Galactic center. 
They suggested that these seven globular clusters come from the Large Magellanic Cloud and have recently been captured by the Galaxy through the Magellanic plane.

As noted above, several previous studies have indicated that M53 and NGC 5053 have originated in a dwarf galaxy that accreted into the Milky Way.
In this respect, the field around the M53 and NGC 5053 is an ideal place to 
search for the tidal tails around the clusters, as well as extra-tidal GC-origin stars that are decoupled from the clusters.
~\citet{Lauchner2006} reported a tidal tail of NGC 5053, and ~\citet{Beccari2008} also suggested a potential tidal tail of M53.
Indeed, ~\citet{Chun2010} detected a tidal-bridge feature between two clusters and tidal common envelope around the clusters, and
~\citet{Jordi2010} also found the extra-tidal substructure around the two clusters.
However, previous extra-tidal studies of the clusters were based on photometric stellar density features in the sky;
they did not investigate the kinematics and chemical properties of the stars in the tidal features with those of clusters.
Therefore, our search for the extra-tidal GC-origin stars from M53 and NGC 5053 is mainly based on the radial velocity and chemical abundance  properties of stars covered by the Apache Point Observatory Galactic Evolution Experiment (APOGEE) survey.
In Section 2, we describe the sample selections for our analysis.
Spatial distributions of cluster member stars and extra-tidal stars are indicated in Section 3. 
Chemical properties of cluster member stars and extra-tidal stars are presented in Section 4.
Finally, the discussion and conclusion are presented in Section 5.

\section{Apogee DATA and Cluster member selection} \label{sec:method}
\begin{figure*}
\includegraphics[width=\textwidth]{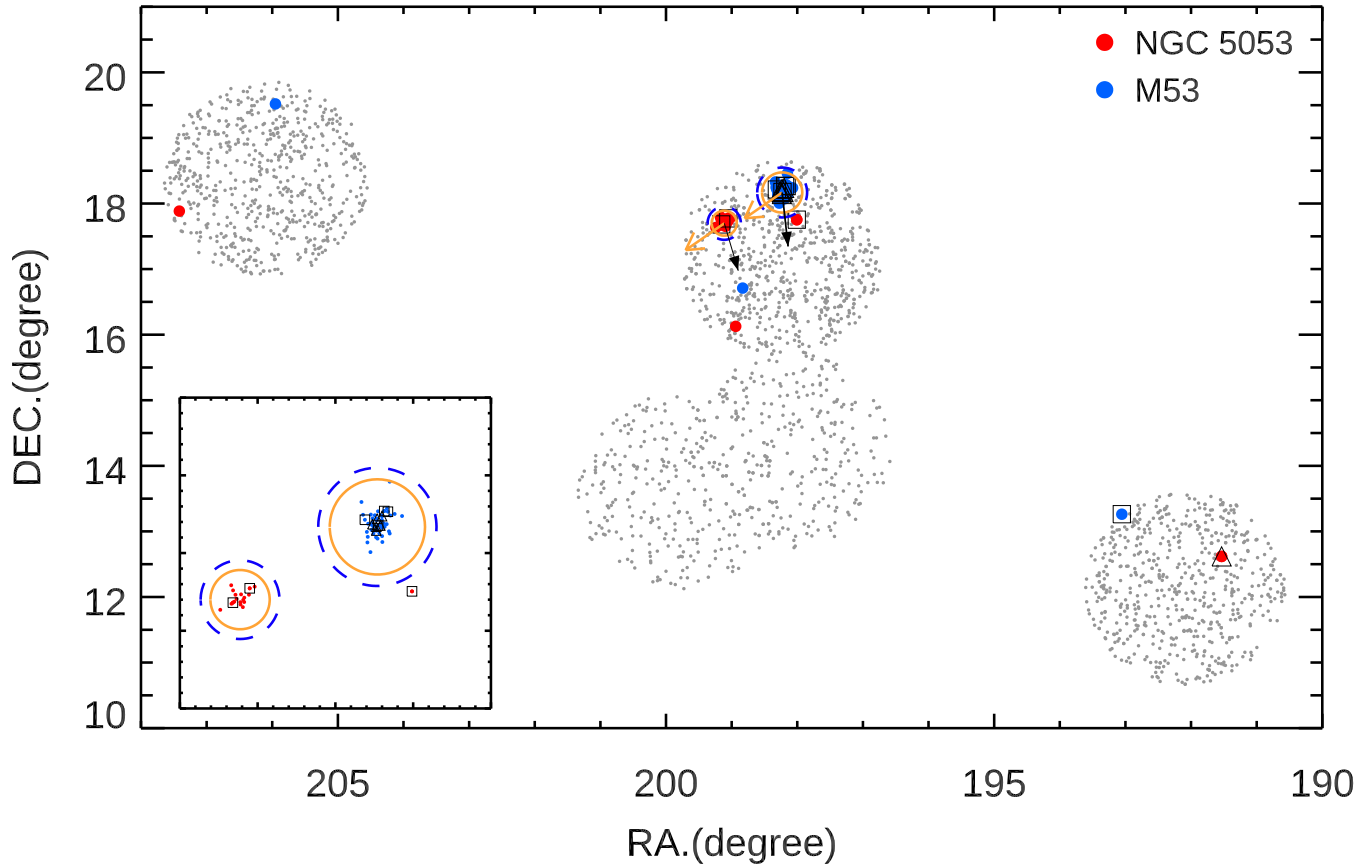}
\caption{Spatial distribution of all stars in the $20\degr \times 10\degr$ field around M53 and NGC 5053.
The subpanel shows the distribution of cluster member candidates inside tidal radii of the clusters.
Tidal radii of two clusters ($18.37'$ for M53 and $11.43'$ for NGC 5053) given by~\citet{Harris1996} (2010 edition) are
indicated by the orange circles. We also present larger tidal radii of $22.8'$ and $15.2'$ which are derived by~\citet{Boer2019} as blue dashed circles.
The direction toward Galactic center was indicated by orange arrow, and the
proper motions of the clusters~\citep{Vasiliev2019} were represented by black arrow.
The points depicted by only blue and red colors are final cluster member candidates in group 1, while
the points with open square and triangle are the stars in group 2 and group 3, respectively (see text).
}\label{fig:dist}
\end{figure*}

The APOGEE survey provides high resolution (R$\sim$22,500)
$H$-band spectra ($\lambda=1.51-1.70 \mu m$). The survey delivers two sets of the stellar parameters and chemical abundances for more
than 20 elements determined by The Cannon~\citep[a data-driven approach to determine stellar parameters and abundances,][]{Ness2015} and
ASPCAP~\citep[APOGEE stellar parameters and Chemical Abundances Pipeline,][]{Garc2016}.
In this work, we use the spectra of APOGEE DR14~\citep{Majew2017} and the data catalogue of The Cannon.
Our initial sample consists of 2,558 redgiant branch (RGB) stars in the $20\degr \times 10\degr$ field around M53 and NGC 5053
that are covered by the APOGEE survey; the distribution of an initial sample of the sky is shown in Figure~\ref{fig:dist}.

Our approach is to search for stars with spectral characteristics that are similar to the cluster members, where we assume that stars that have decoupled from the clusters will show similar properties in chemical abundances and kinematics to the member stars of the clusters. To measure spectral similarity, we apply t-distributed stochastic neighbor embedding (t-SNE) algorithm to the datasets of chemical abundances and radial velocities.
t-SNE is a machine-learning algorithm for visualization with clustering similar features together of the high-dimension data
into lower-dimension.
It calculates the probability distribution of similarities for each pair of points in the high and low dimensional spaces, respectively, and then
tries to find locations in the lower-dimensional space to minimize the difference between these probability distributions (or similarities)
for an optimal representation of data points in lower-dimensional space. 
The Kullback-Leibler (KL) divergence~\citep{Kullback1951}, a measure of direct divergence between two probability distributions of overall data points using a gradient descent method, is utilized to measure the difference of probability distributions.
A detailed explanation of t-SNE algorithm is described in~\citet{Maaten2008} and~\citet{Pezzotti2015} (see their papers for more details). 
The t-SNE algorithm is now extensively used in astronomy as a classification algorithm~\citep{Lochner2016, Matij2017, Valentini2017, Traven2017}.
In particular, ~\citet{Anders2018} and~\citet{Kos2018} demonstrated the capability of t-SNE to identify stellar populations 
in the Milky Way by recognizing the clusters and related field stars, as well as chemically peculiar stars.

\begin{figure}
\includegraphics[width=\columnwidth]{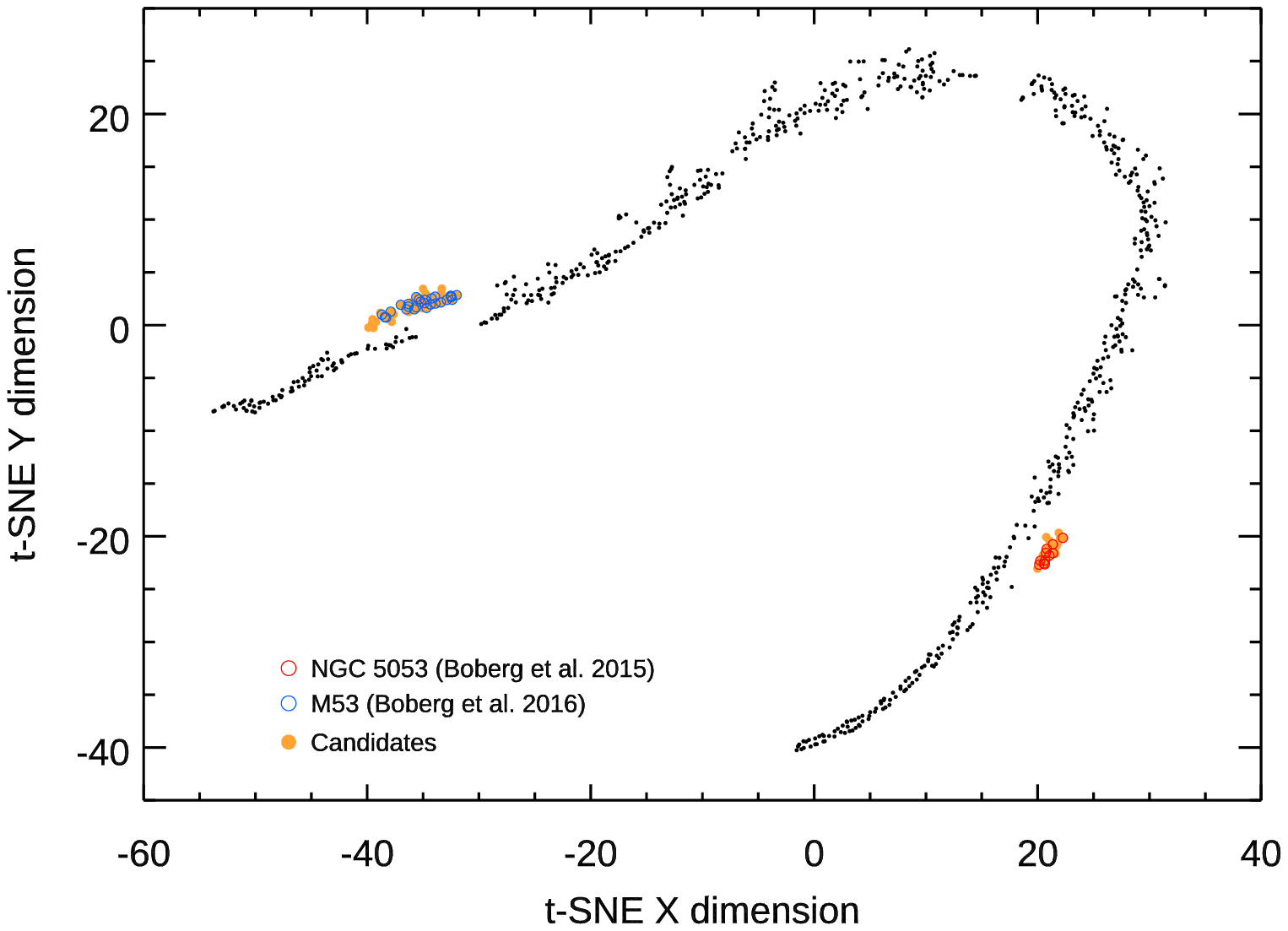}
\caption{t-SNE projection of stars around M53 and NGC 5053. The points in two subgroups (orange points) at -40 and 20 on the t-SNE X 
dimension axis are selected cluster member candidates. M53 and NGC 5053 member stars confirmed by~\citet{Boberg2015,Boberg2016}
are indicated by blue and red open circles, respectively. }\label{fig:tsne}
\end{figure}

We use the t-SNE algorithm included in the scikit-learn python package.
As input data for t-SNE, the radial velocity and chemical abundances of The Cannon were 
used.
As different chemical elements have different degrees of uncertainties, 
we tested several combinations of atomic elements to find the optimal set of elements for our t-SNE analysis. Out of 20 elemental abundances of The Cannon, our analysis is based on 19 elements; Na is excluded because this element is difficult to measure in near-infrared spectra for 
metal-poor stars.
Any stars with poor stellar parameters are not included in the t-SNE computation. Instead, a separate selection process
is applied to these stars (described below).
Figure~\ref{fig:tsne} shows the t-SNE projection for the stars around M53 and NGC 5053 in Figure~\ref{fig:dist}.
The t-SNE projection shows the groups with similar radial velocity and chemical abundances together.
One can easily identify that there are two well-defined isolated groups (i.e., the orange dots). It turns out that the group at t-SNE with an X-value of -40 is mostly composed of member stars of M53, while the other group at t-SNE with an X-value of 20 is mostly composed of member stars of NGC 5053.
The cluster member stars confirmed by~\citet{Boberg2015,Boberg2016} are also well distributed in the two isolated groups.
Therefore, we consider all the stars in the two isolated groups as being candidate members of M53 and NGC 5053.

\begin{figure}
\includegraphics[width=0.9\columnwidth]{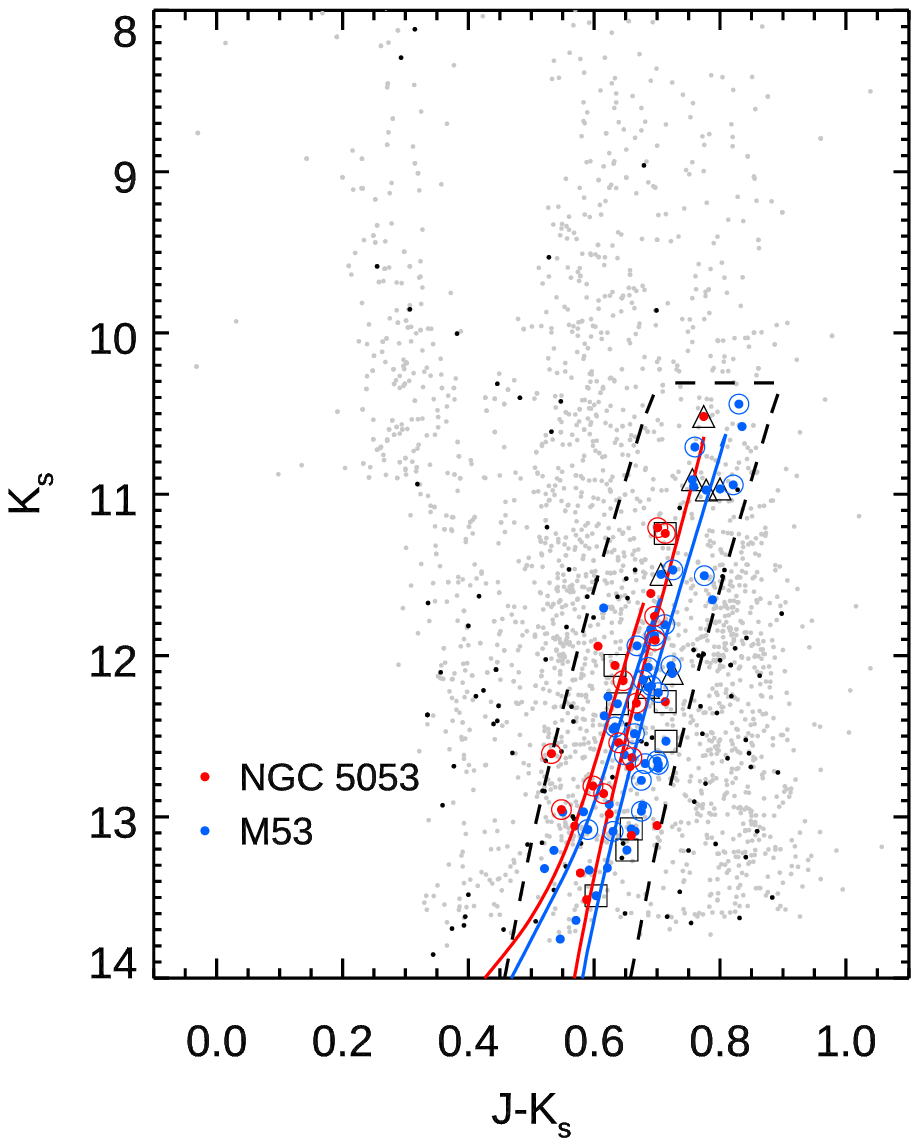}
\caption{$(J-K_s, K_s)$ CMD for the RGB stars around M53 and NGC 5053. The gray dots indicate the all stars in the Figure~\ref{fig:dist}, and black dots are selected member stars from the t-SNE algorithm and those based on the radial velocity criteria (see text).
The MIST isochrones with $\mathrm{[Fe/H]=-2.27}$ and $-2.10$ and age of 13 Gyr were plotted as the red and blue solid lines, respectively.
The dashed lines are the color and magnitude boundaries for filtering the field stars.
Final cluster member candidates in group 1, group 2 and group 3 are 
represented as Figure~\ref{fig:dist}.
The open circles are confirmed cluster member stars recognized in~\citet{Boberg2015,Boberg2016}.
}\label{fig:cmd}
\end{figure}

\begin{figure}
\includegraphics[width=\columnwidth]{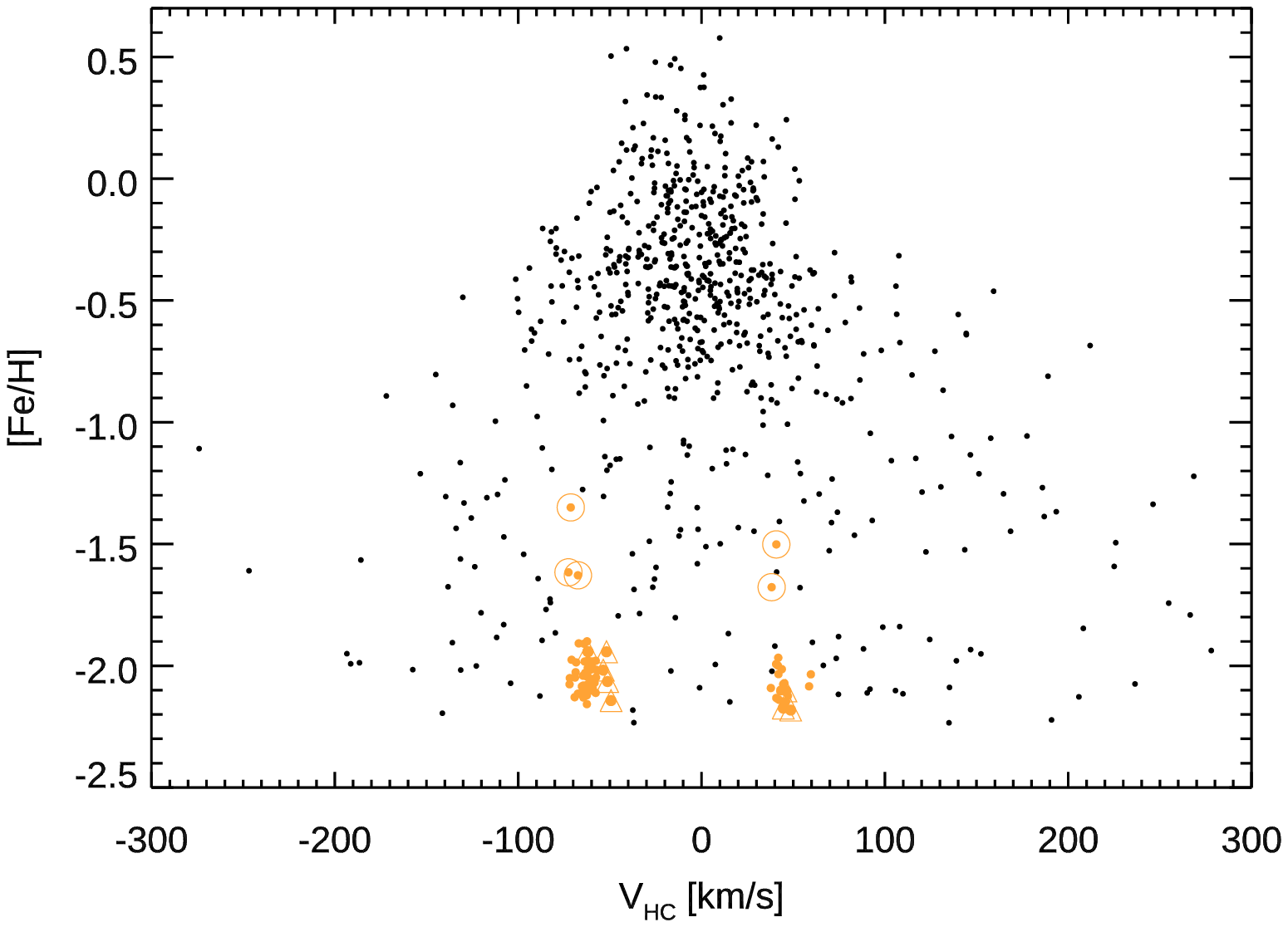}
\caption{Heliocentric radial velocities versus metallicities of the stars around M53 and NGC 5053 in Figure~\ref{fig:dist}. 
The cluster member candidates in t-SNE or CMD filtering analysis are indicated by orange dots. {\bf The stars in group 2 are indicated by open
triangle.} Note that five metal-rich stars (orange open circles) were excluded from the final list of the candidates (see text).
}\label{fig:metal}
\end{figure}

In addition to the t-SNE algorithm, we apply separate selection processes. The additional selection processes are used to refine the t-SNE selection and to search for additional member candidates from the sample where we could not apply the t-SNE process due to various reasons (e.g., inaccurate stellar parameters). More specifically, we filter the sample based on the radial velocities, the location in the color-magnitude diagram (CMD), and metallicities $\mathrm{[Fe/H]}$. These filtering processes are guided by properties of known cluster members and the t-SNE selected candidates.

For the baseline criteria, we select stars with a radial velocity range of about $\pm15$ km s$^{-1}$ from the mean value of the two
clusters, $V_{rad}=44$ km s$^{-1}$ for NGC 5053~\citep[][adopted from ~\citealp{Harris1996}]{Pryor1991,Geisler1995,Yan1996} and $V_{rad}=-63$ km s$^{-1}$ for M53~\citep[][adopted from ~\citealp{Harris1996}]{Lane2010}.
We further apply filtering based on their location of the CMD.
Figure~\ref{fig:cmd} shows the $(J-Ks, Ks)$ CMD of all stars (gray dots) in Figure~\ref{fig:dist} and the stars (black dots)
filtered by the radial velocities criteria.
The filtering criteria for the CMD (dashed lines in Figure~\ref{fig:cmd}) is determined by considering the theoretical isochrones of the clusters and the distribution of confirmed and t-SNE selected cluster members. The isochrones are derived from MIST~\citep{Choi2016,Dotter2016} for $\mathrm{[Fe/H]=-2.10}$ and  $-2.27$ and age of 13 Gyr that are relevant to M53 and NGC 5053.
With respect to these isochrones, we define our selection boundary with the color-width of about 0.1 mag and a magnitude range of $K_s=10.3\sim14.0$, which comfortably  enclose all the confirmed cluster members~\citep{Boberg2015,Boberg2016}. We note that, among the t-SNE selected candidates, three stars were outside of this boundary, and they are dropped from our candidate sample. 

Finally, we apply criteria for $\mathrm{[Fe/H]}$. 
Figure~\ref{fig:metal} shows the distributions of $\mathrm{[Fe/H]}$ and the heliocentric radial velocity for our initial sample, together with the member candidates selected from the t-SNE or CMD filtering process.\footnote{We note that some stars that are very close to the domain of member candidates are not selected as member candidates because they are not in the selection boundary in the CMD of Figure~\ref{fig:cmd}.} 
In addition to radial velocity, the member candidates show reasonable metallicity distribution that is consistent with the metallicities of clusters from previous studies. 
We also find a few member candidates show relatively a high value of metallicity greater than $\mathrm{[Fe/H]=-1.8}$. While we are inclined to drop these stars of high metallicity, we want to be careful, as several studies based on the APOGEE catalogue have reported possible systematic bias in their metallicity at a low metallicity ($\mathrm{[Fe/H]<-1.0}$) domain. 
As a cautionary measure, we conduct our own spectral synthesis analysis to independently derive metallicity and chemical abundances (details are described in Section 4). We assign metallicity range for cluster member candidates with $-2.3<\mathrm{[Fe/H]}<-1.8$ for M53; and $-2.45<\mathrm{[Fe/H]}<-1.85$ for NGC 5053. These are determined from the mean metallicities and the standard deviations from our own analysis.
The candidate stars of high metallicity from the APOGEE catalog resulted in similarly high metallicity, based on our analysis.
Thus, the five stars with high metallicity are removed from the cluster candidates. 


\begin{deluxetable}{ccccccc}
\tabletypesize{\normalsize}
\tablewidth{0pt}
\tablecaption{Number of selected cluster member (and extra tidal star) candidates and selection method \label{tab:table1}}
\tablehead{
\colhead{Group} & \colhead{M53} & \colhead{NGC 5053} & \multicolumn{4}{c}{Selection method} \\
\colhead{}   & \colhead{}   &   \colhead{}   & \colhead{t-SNE}        &   \colhead{CMD}   &  \colhead{$V_{HC}$}  &  \colhead{[Fe/H]} 
}
\startdata
Group 1	& 40 (2)	& 18 (2)	& O 	& O 	& O	& O \\ 
Group 2	& 5 (1)	& 3 (1)	& X 	& O 	& O	& O \\ 
Group 3	& 6 (0)	& 1 (1)	& X 	& O 	& O	& X  \\
\hline
total		& 51	(3)	& 22	(4)	& - 	& - 	& -	& -  \\
\enddata
\end{deluxetable}

\begin{deluxetable}{rccc}
\tabletypesize{\normalsize}
\tablewidth{0pt}
\tablecaption{Selection creteria\label{tab:table2}}
\tablehead{
\colhead{Method} & \colhead{M53} & \colhead{NGC 5053} }
\startdata
t-SNE (X)				& -40			& 20				\\
CMD ($J-K_{s}$)		& 0.45 $\sim$ 0.90	& 0.45 $\sim$ 0.90	\\
~~~~~~~~~~$(K_{s})$	& 10.3 $\sim$ 14.0	& 10.3 $\sim$ 14.0	\\
$V_{HC}$ (km/s)		& -78.2 $\sim$ -48.2	& 30.1 $\sim$ 60.1	\\ 
$[$Fe/H$]$ (dex)		& -2.3 $\sim$ -1.8	& -2.45 $\sim$ -1.85	\\
\enddata
\end{deluxetable}

We finally identify $73$ stars (51 for M53 and 22 for NGC 5053) as being cluster member candidates.
Note that 33 and 13 stars are cross-matched with sample stars of~\citet{Boberg2015,Boberg2016} for M53 and NGC 5053, respectively.
We classify them into three groups based on the likelihood of membership. All the candidates conform to our CMD criteria. 
Group 1 represents candidates that are most likely selected via t-SNE and also conform to our radial velocity and metallicity criteria. 
Group 2 represents candidates that conform to the radial velocity and metallicity criteria, but for which t-SNE was not applicable. 
Group 3 are candidates filtered by radial velocity criteria only, as neither t-SNE nor metallicity cut is applicable given the poor spectral quality.
The member candidates in Group 1, Group 2 and Group 3 are represented as colored points, open squares, and open triangles in Figure~\ref{fig:dist} and Figure~\ref{fig:cmd}.
In Table~\ref{tab:table1}, we summarize the number of cluster member stars in each group and indicate applied selection methods. 
The number in parentheses indicates the number of extra-tidal stars. In Table~\ref{tab:table2}, the selection criteria for each selection
method are indicated.


\section{Spatial distribution and extra-tidal stars from the clusters} \label{sec:analysis}
In Figure~\ref{fig:dist}, we show the spatial distribution of our 73 cluster member candidates in the sky.
Member candidates of M53 are marked with blue points, and those of NGC 5053 are marked with red. Also shown
are the tidal radii of each cluster.
In this study, we derive the tidal radii of M53 and NGC 5053 using the structural parameters (core radius and central concentration) in~\citet[][2010 edition]{Harris1996}, which are actually derived by the King model fitting to the radial density profiles of clusters~\citep{Lehmann1997}.
The calculated tidal radii are $18.37'$ for M53 and $11.43'$ for NGC 5053.
We are certain that the derived tidal radii are appropriate for describing the limit of the clusters. 
The radial density profiles of M53 and NGC 5053 in~\citet{Chun2010} showed that overdensity features with clear slope changes in the profile that
depart from the King model start at $15'$ and $10'$, respectively, and extend to $34'$ from the cluster center. 
The tidal radius is entirely dependent on the fitting model, and indeed larger tidal radii of  $22.8'$ and $15.2'$ for M53 and NGC 5053, respectively,
were reported by~\citet{Boer2019}. 
Using the data of Gaia DR2, they fitted a Spherical Potential Escapers Stitched (SPES) model which shows
a more detailed description of stars at the escape energy to the density profile of the clusters.
We note that adopting larger tidal radii does not change the final results, because identified extra-tidal stars are still beyond
the larger tidal radii. In Figure~\ref{fig:dist}, we present the larger tidal radii by~\citet{Boer2019} as dashed blue circles.

We see most of the candidates are indeed located within the tidal radii of the clusters.
We do not see any obvious tidal extension of candidates, but given the small number of samples, we do not expect to see one, even if there is one
present. It is apparent, however, that several member candidates are located well beyond the tidal radii. 
We find a total of seven possible extra-tidal stars; three stars out of the 51 candidates of M53 members, and four out of the 22 of NGC 5053.
Among seven likely extra-tidal stars, four are from Group 1, two are from Group 2, and one is from Group 3 (see Table~\ref{tab:table1}).
We note that ~\citet{Tang2018} recently investigated the stars of NGC 5053 using the same APOGEE data that we used,
but they were not able to identify any extra-tidal star. They only considered the stars within three times the tidal radius of NGC 5053, 
and all our candidates are located beyond their search radius.
The most distant extra-tidal stars of M53 and NGC 5053 are more than 8 degree away from the cluster.
At this large scale, our initial search sample is strongly biased by the coverage of the APOGEE survey, which is obvious from Figure~\ref{fig:dist}. Given a small number of extra-tidal candidates and very small coverage of the initial samples, we are limited in investigating the spatial distribution of extra-tidal stars. However, in the vicinity of two clusters within a few degrees from the clusters, there are some notable aspects in the distribution.

\begin{figure*}
\plottwo{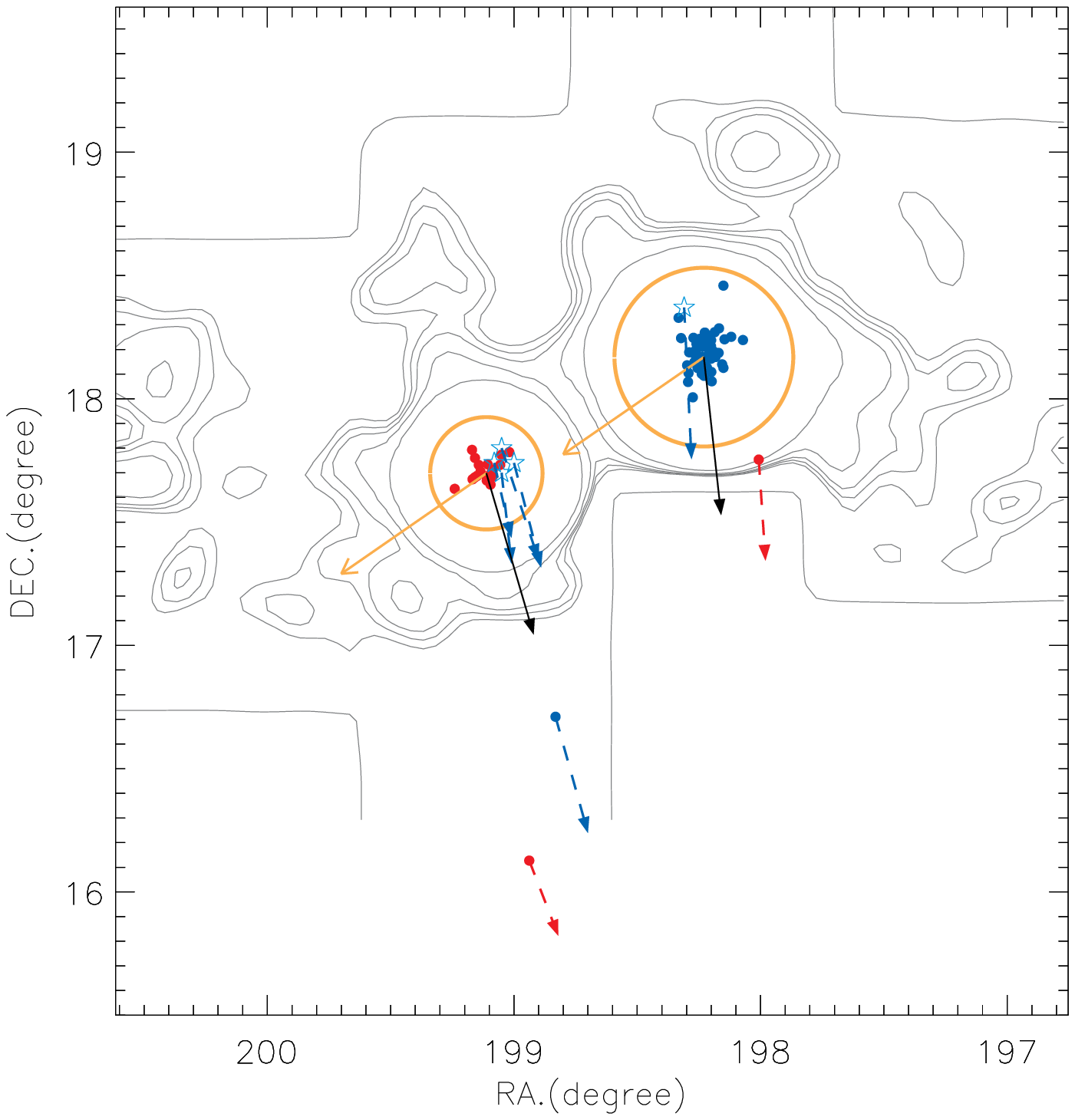}{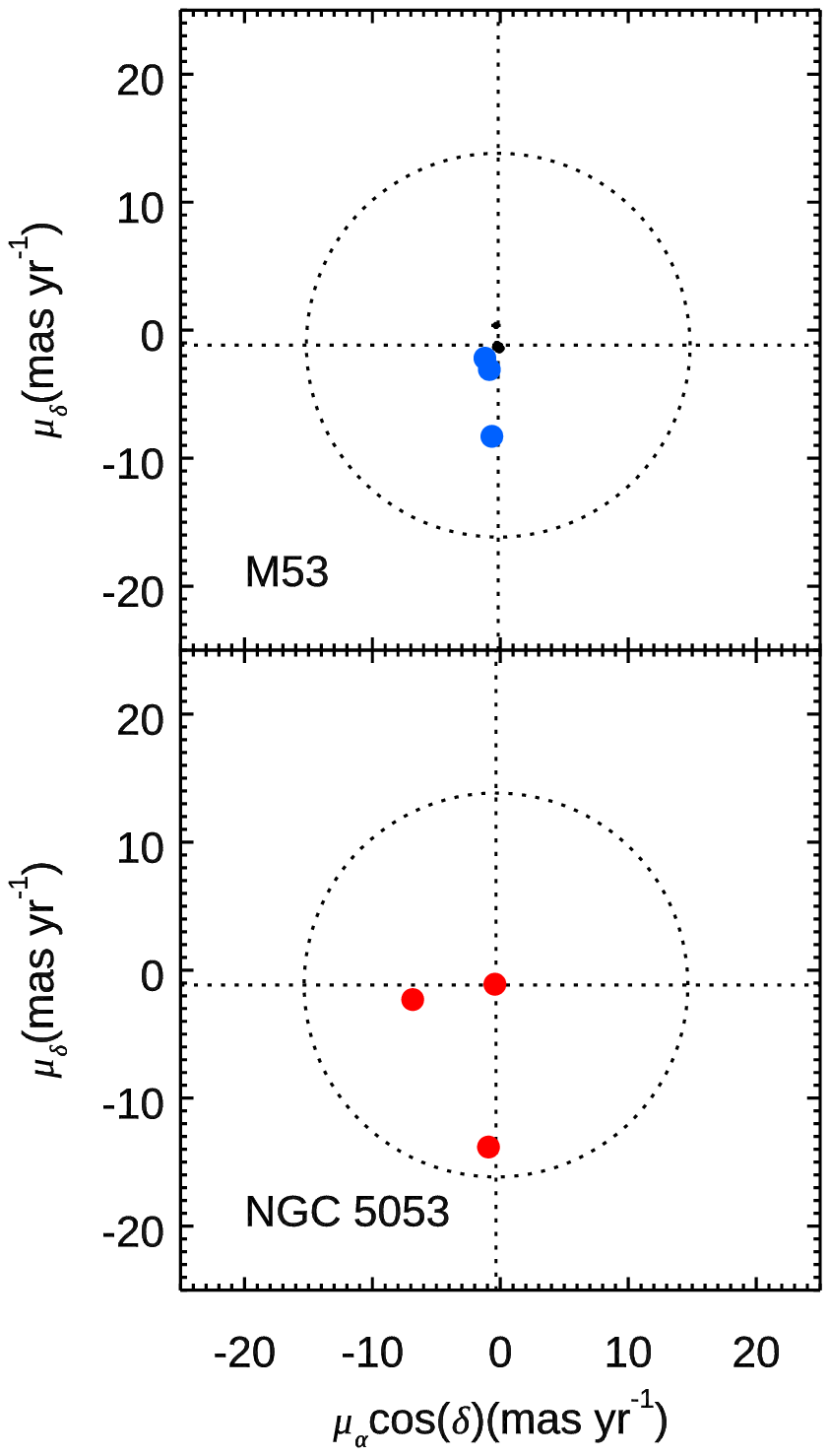}
\caption{\textit{Left:} Expansion of spatial distribution of stars near the two clusters with isodensity contour map of~\citet{Chun2010}. The tidal radii of clusters are indicated by orange circles. The direction toward the Galactic
center and proper motion of the clusters are represented by orange and black arrows, respectively. 
The dotted arrows are the proper motion~\citep[Gaia DR2]{Gaia2018} direction for extra-tidal stars. 
The blue points are for cluster member candidates of M53 and red points are for NGC 5053.
Five extra-tidal RR Lyrae stars of M53 from~\citet{Kundu2019} are
also plotted by open stars. 
\textit{Right:} Proper motion diagram of extra-tidal stars. The intersection of dotted line is nominal proper motion of the 
cluster~\citep{Vasiliev2019}. 
}\label{fig:contour}
\end{figure*}

The left panel of Figure~\ref{fig:contour} shows the zoomed-in view of the spatial distribution centered around the two
clusters. Overlaid are the stellar density contours around M53 and NGC 5053 from~\citet{Chun2010}. 
In addition to the member candidates from this work, we overplot the extra-tidal RR Lyrae stars of M53 of~\citet{Kundu2019}.
The solid black arrows indicate the proper motion of M53 and NGC 5053. 
The proper motions of individual extra-tidal star candidates, if available from the Gaia DATA Release 2, are indicated by dotted arrows.
They all show similar proper motion to their suggested parental clusters. The right panel of Figure~\ref{fig:contour} shows 
the proper motion diagram of six stars in our seven extra-tidal stars, and the proper motions are consistent with those of 
the suggested parental clusters (within about $15$ mas yr$^{-1}$, dotted circle). 
It is also notable that three extra-tidal star candidates from this work are located along the leading direction of the cluster proper motion.

The stellar density contour of~\citet{Chun2010}, and their possible association with extra-tidal candidates can be interesting. 
However, the stellar density contours of~\citet{Chun2010} are limited in their coverage; thus,
it is not trivial to associate the features in the density contours to the location of extra-tidal candidates.
Even with these limitations, it is notable that there are small clumps along the trailing direction of the cluster proper motion and 
on the extension line between the extra-tidal candidates and the clusters.
Marginal density contours also seem to approach or bend toward extra-tidal stars.
We note that tidal tails or extra-tidal stars should be aligned with the cluster’s orbit~\citep{Combes1999,Dehnen2004,Jordi2010,Eyre2011},
but extra-tidal stars located in different positions from the cluster do not need to have
similar proper motion~\citep[e.g.,][]{Anguiano2016}.
Therefore, several properties of extra-tidal stars, such as the alignment toward the leading direction of the cluster proper motion, similar proper motions, and marginal association with stellar density contour, support the idea that these stars are extra-tidal stars decoupled from two globular clusters.

Two other interesting features in stellar density contour of~\citet{Chun2010} are the tidal bridge feature and the tidal common envelope between the two clusters. 
~\citet{Chun2010} suggested dynamical interaction between clusters. 
In this regard, it is interesting to note that one of the extra-tidal stars of NGC 5053 $(\alpha, \delta) \sim (198.0, 17.8)$ is indeed located within the M53 side of the common envelope,
which raises the interesting possibility that this star was originally a member of NGC 5053, was stripped from its initial prenatal cluster, and is now under the gravitational influence of M53. 
~\citet{Kundu2019} reported five extra-tidal RR Lyrae stars of M53 from Gaia DR2, and we find that four of them are located 
inside the tidal radius of NGC 5053 and have very similar proper motion to that of M53.
This may provide another piece of evidence that the two clusters are dynamically interacting and possibly swapping their member stars. 
However, we are concerned that the four extra-tidal RR Lyrae stars of~\citet{Kundu2019} are simply member stars of NGC 5053.
Since M53 and NGC 5053 show similar proper motion and have similar apparent magnitude, accidentally misidentification of the member stars of NGC 5053 
as extra-tidal sources of M53 is sufficiently possible.
Indeed,~\citet{Ngeow2020} recently reported that the four extra-tidal RR Lyrae stars identified by~\citet{Kundu2019} were already-known RR Lyrae stars of NGC 5053
in the ``Updated Catalog of Variable Stars in Globular Clusters''~\citep{Clement2001,Clement2017}.
Therefore, we consider these RR Lyrae stars as being the member stars of NGC 5053, not extra-tidal stars of M53. Note that the radial velocity of these stars and
the period-luminosity-metallicity relation of RR Lyrae in the two clusters are helpful for further discussion.

\begin{figure*}
\plottwo{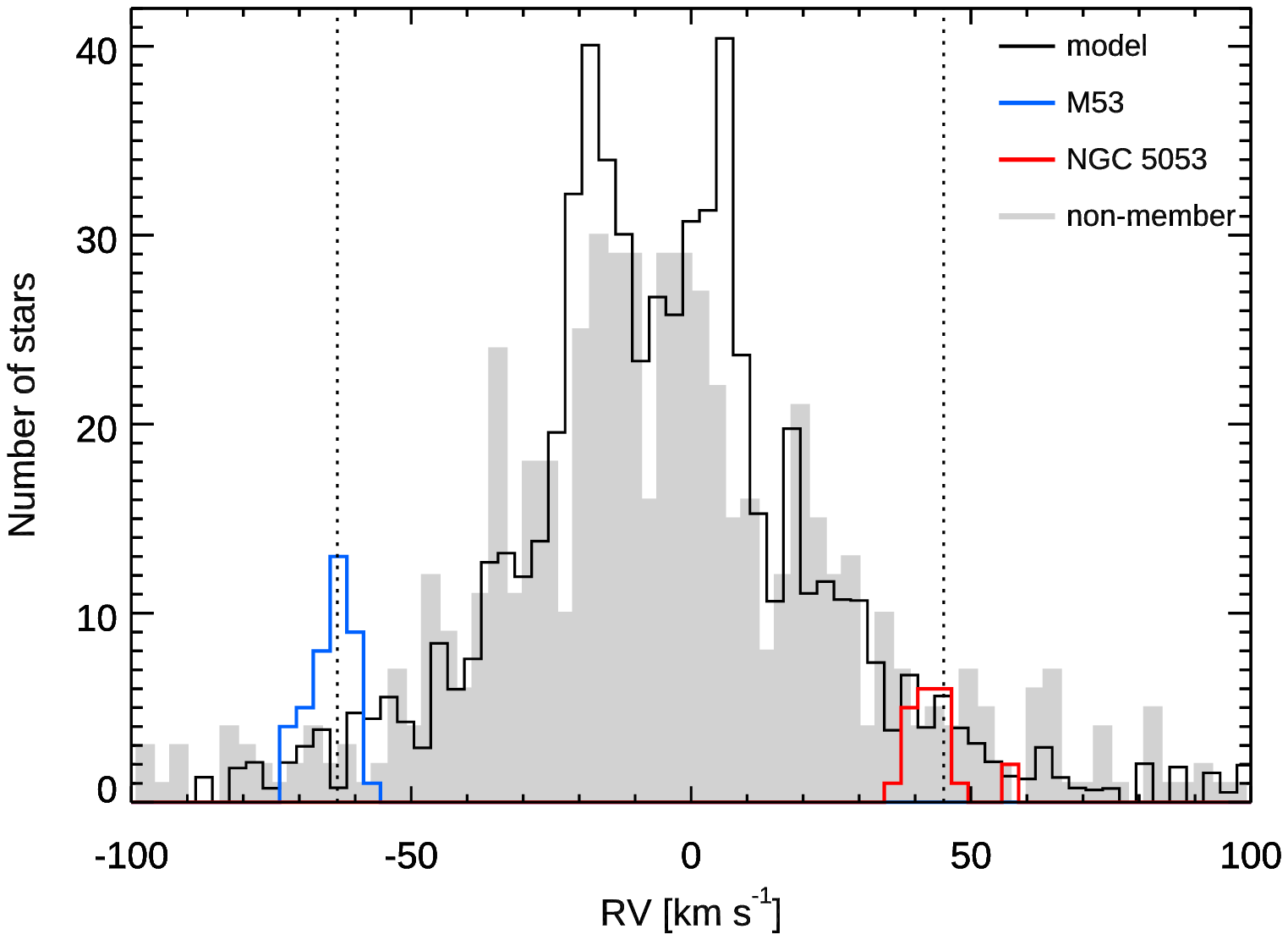}{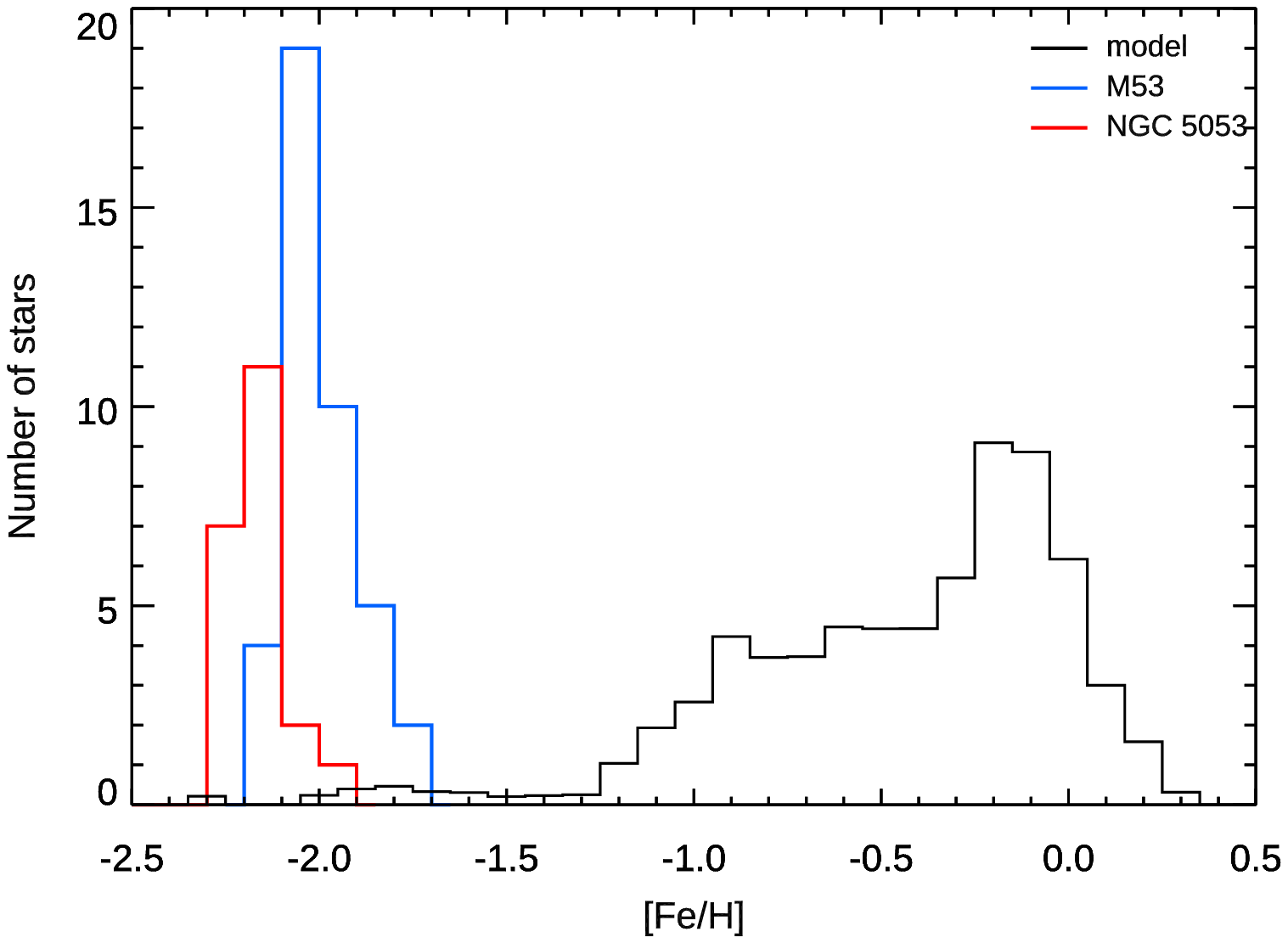}
\caption{\textit{Left:} Radial velocity distribution for observed stars and model populations. Blue and red histograms are
radial velocity distribution of M53 and NGC 5053, respectively. The dotted lines are nominal radial velocities of the clusters.
The observed cluster non-member are overplotted by grey histogram.
\textit{Right:} Metallicity
distribution of cluster member stars and model populations }
\label{fig:distribution}
\end{figure*}

To investigate possible contamination by field stars to our sample of cluster member candidates,
we use the Besan\c{c}on Galaxy model~\citep{Robin2003} to simulate the radial velocity and metallicity distribution of field stars.
We generate 100 Galaxy models covering the same areas around the clusters.
We then apply the same selection criteria we used for the membership selection in the CMD space (i.e., the dashed lines in Figure~\ref{fig:cmd}).
Figure~\ref{fig:distribution} shows the radial velocity (left panel) and metallicity (right panel) distribution of the simulated stars.
Radial velocity distribution is normalized to the total number of observed non-member stars, while the metallicity distribution is normalized to the total number of cluster member stars. The distributions for the cluster members are indicated by the blue (M53) and red (NGC 5053) histograms. 
We note that the metallicities of the clusters in Figure~\ref{fig:distribution} are from our own analysis, which will be described in Section 4.

In the radial velocity distribution, we find that the mean velocities of the stars designated as cluster members
are $-63.2\pm4.1$ km s$^{-1}$ for M53, and $45.1\pm5.3$ km s$^{-1}$ for NGC 5053.
These mean radial velocities are in agreement with the previously derived values of $-63.2\pm0.5$ km s$^{-1}$
for M53~\citep{Boberg2016}, and $42.0\pm1.4$ km s$^{-1}$ for NGC 5053~\citep{Boberg2015}.
Unfortunately, it is apparent that the radial velocity distributions of cluster stars are not much different from the Galaxy model distribution.
M53 distribution shows a peak at its mean velocity, but this does not seem to be dominant, compared to the model distribution.
A Kolmogorov-Smirnov (KS) test for the null hypothesis that observed radial distributions of the clusters and the model come from the same distribution also provides $p$-values of about $P=0.07$ for M53, and $P=0.37$ for NGC 5053; this indicates that we cannot
reject the null hypothesis.
On the other hand, metallicity distributions of the clusters and the Galaxy model show an interesting feature.
The Galaxy model predicts fewer than one star of such low metallicity ($\mathrm{[Fe/H]\sim-2.0}$) in the cluster field.
However, the metallicity distributions of two clusters show apparent and prominent peaks at low metallicities. 
The $p$-values of the KS test to compare the cluster and model metallicities are also almost zero ($7.64\times10^{-34}$ for M53, and $1.06\times10^{-18}$ for NGC 5053). 
It is unlikely that the observed and expected metallicity distributions come from the same parent distribution. 
In addition, we search for the stars from the Galaxy model that are consistent with our criteria for member candidates (i.e., based on metallicity and radial velocity criteria), 
and we find that only two field stars are in this condition.
Based on a comparison with the Galaxy model, the radial velocities of the identified cluster and extra-tidal stars are consistent with 
those predicted by the Galaxy model, even though our sample stars are clearly different populations from the Galaxy populations in terms of metallicity.
There are only two halo interlopers, which are consistent with both the radial and the metallicity criteria of the member candidates.
Therefore, our extra-tidal stars are likely associated with the two globular clusters, even though we cannot definitively
exclude the possibility that they are normal field stars.

\section{Chemical properties of clusters and extra-tidal stars} \label{sec:chemical}
The anomalies of light-elements, such as C, N, O, F, Na, Al, Mg and Si, are unique features that are found only in globular clusters; thus,
the Galactic field stars with light-element patterns similar to those seen in
second-populations of the globular clusters are considered to be escaped stars from the globular clusters.
Indeed, several studies have reported the Galactic field stars with GC-like abundance patterns and discussed the association thereof with globular clusters~\citep[e.g.,][]{Ramirez2012,Wylie2012,Carretta2013,Martell2016,Schi2017b}.
Therefore, if our extra-tidal candidate stars show GC-like abundance patterns, 
a physical association with the clusters will be strongly supported.
Many previous studies~\citep{Meszaros2015,Jonsson2018,Tang2018,Masseron2019} have suggested that 
the chemical abundances for metal-poor stars ($\mathrm{[Fe/H]}<-1.0$) provided by the APOGEE pipeline may exhibit systematic offset.
Thus, we reanalyze the APOGEE spectra of the cluster member candidates and manually estimate their chemical abundances.
We focus on atomic elements Mg, Al, and Si, because their spectral lines are relatively prominent in the APOGEE spectra for metal-poor stars. 
We investigate abundance anomalies for these elements, as well as the chemical association of extra-tidal stars with the clusters.

\subsection{Stellar parameters and synthetic fitting}
To estimate the chemical abundances for the clusters and extra-tidal stars, we photometrically calculate the stellar atmospheric parameters. 
Using the relations of~\citet{Gonz2009}, the effective temperatures ($T_{\mathrm{eff}}$) are derived from broadband $B, V$~\citep{Boberg2015,Boberg2016} and 2MASS
$J, H, K$~\citep{Skrut2006} photometries.
The reddening correction is applied with $E(B-V)=0.018$ for M53 and $E(B-V)=0.015$ for NGC 5053~\citep{Schla2011}.
The values of $T_{eff}$ obtained from $B-V$, $V-J$, and $J-K$ are averaged and adopted as final $T_{\mathrm{eff}}$, 
or those from $J-K$ are chosen for the stars with only $J-K$ color.
The surface gravities are calculated from the following relation:
\begin{equation}
\log{g} = \log{g_{\odot}} + \log\left(\frac{M_{\star}}{M_{\odot}}\right) + 4 \log\left(\frac{T_{\mathrm{eff\star}}}{T_{\mathrm{eff\odot}}}\right) + 0.4(M_{bol,\star}-M_{bol,\odot}).
\end{equation}
The solar values of $\mathrm{log}$ $g_\odot=4.438$, $M_{bol.\odot}=4.75$, and $T_{eff,\odot}=5772$ K ~\citep{Prsa2016} are used, and
we assume a mass of $0.8M_\odot$ for our sample stars. The bolometric correction (BC) values are estimated from the relation 
between BC values and $J-K$ color of~\citet{Monte1998}. 
Distance moduli of $(m-M)_V=16.32$ for M53, and $(m-M)_V=16.23$ for NGC 5053
~\citep[][adopted from~\citealp{Harris1996}]{Kopacki2000,Arellano2010} are used.
The equation $v_t=\mathrm{2.24-0.3\times log}$  $g$ from~\citet{Meszaros2015} is then used to calculate micro-turbulence velocity ($v_t$).
Table~\ref{tab:table3} shows the sample of derived stellar parameters.

\begin{table*}
\begin{center}
\caption{Stellar parameters, metallicity, and abundances of M53 and NGC 5053}
\label{tab:table3}
\begin{tabular}{ccccccccccccc} 
\hline \hline  
2MASS ID & $T_{\mathrm{eff}}$ & log $g$ & $v_t$ & [Fe/H]  & $\sigma_{[Fe/H]}$ & [Mg/Fe] &  $\sigma_{[Mg/Fe]}$ & [Al/Fe] & $\sigma_{[Al/Fe]}$ & [Si/Fe] & $\sigma_{[Si/Fe]}$ & Cluster\\ 
\hline 
2M13434835+1931084  & 4836 &  1.10  &  1.91   &  -2.051  &  0.020   &  0.500  &   0.025  &  -0.029   &  0.060   &  0.393   &  0.072 & M53 \\
2M13151955+1642373  & 4656 &  1.36  &  1.83   &  -2.223  &  0.088   &  0.482  &   0.096  &  -0.138   &  0.090   &  0.449   &  0.090 & M53 \\
2M13123617+1827323  & 4634 &  1.58  &  1.77   &  -2.013  &  0.019   &  0.512  &   0.019  &  -0.037   &  0.029   &  0.509   &  0.027 & M53 \\
2M13124987+1811487  & 4647 &  1.39  &  1.82   &  -1.908  &  0.100   &  0.289   &   0.106 &  -0.419   &  0.105   &   0.321  &   0.114 & M53 \\
2M13124768+1810060  & 4421 &  0.96  &  1.95   &  -2.012  &  0.080   &  0.310  &   0.105  &   1.101   &  0.085   &  0.482   &   0.083 & M53 \\
2M13130945+1811188   & 4678 &  1.36  & 1.83    & -1.985  &  0.025   &   0.439  &   0.059  &   0.801   &  0.039   &  0.495   &  0.036 & M53 \\
2M13121714+1814178  & 4558 &  1.48  &  1.79   &  -1.972  &  0.075  &   0.393  &   0.084   &  -0.135  &  0.083   &  0.343   &  0.076 & M53 \\
2M13124082+1811099  & 4672 &  1.64  &  1.75   &   -2.084  & 0.016  &   0.512   &  0.017   &  0.993   &  0.052   &   0.600  &   0.038 & M53 \\
...                                    &  ...     &  ...      &  ...       &    ...        & ...        & ...            &    ...      &    ...        &    ...       &     ...      &     ...     & ...     \\
2M13120179+1745121  & 4769 &  1.27  &  1.86   &   -2.182  & 0.044  &   0.505   &  0.054   &  -0.073  &   0.046  &   0.559  &   0.047 & NGC 5053 \\
2M13154512+1607370  & 4799 &  1.64  &  1.75   &   -2.141 &  0.123  &   0.405   &  0.126   &  0.199   &  0.159   &  0.483   &   0.130 & NGC 5053 \\
2M13493976+1753033  & 4861 & 1.23  &  1.87   &   -2.103  & 0.054  &   0.257  &   0.114  &  -0.003   &  0.114   &  0.496   &  0.106  &  NGC 5053 \\
2M13161223+1746228  & 4464 & 0.91  &  1.97   &   -2.174  & 0.016  &   0.271   &  0.016   &  0.270   &  0.101   &  0.291  &   0.053  &  NGC 5053 \\
2M13160457+1747017  & 4684 & 1.69  &  1.73   &   -2.322  & 0.072   &  0.223    &  0.088  &  0.942    & 0.123   & 0.442   &   0.073  & NGC 5053 \\
2M13162073+1741059  & 4738 & 1.51  &  1.79   &   -2.109   & 0.021  &  0.067   &  0.077   &  0.907   &  0.102   &  0.395   &  0.077 & NGC 5053 \\
2M13162226+1741536  & 4850 & 1.42  &  1.81   &   -2.193   & 0.029  &   0.219  &   0.052  &  1.038   &  0.104  &   0.648  &  0.030 & NGC 5053 \\
2M13162059+1742464  & 4560 & 1.14  &  1.90   &   -2.196   & 0.032  &  0.328    & 0.084   &  0.040    & 0.105   &  0.268   & 0.087 & NGC 5053 \\
...                                    &  ...     &  ...      &  ...       &    ...        & ...        & ...            &    ...      &    ...        &    ...       &     ...      &     ...     & ...     \\
\hline
\end{tabular}
\end{center}
\end{table*}

Based on the derived stellar parameters, we estimate chemical abundances of individual elements (i.e., Fe, Mg, Al, and Si) by synthetic spectral fitting 
to the interesting atomic lines in the observed spectra. 
The synthetic spectra are generated by Turbospectrum~\citep{Alvarez1998,Plez2012} with 
the atmospheric models interpolated from MARCS model grid~\citep{Gustaf2008}, 
and the internal APOGEE DR14 atomic/molecular linelist (linelist 20150714) is used in the model calculation.
The calculated synthetic spectra are then convolved by a line-spread function (LSF) that is used in ASPCAP to match the
observed line profile and the spectral resolution. 
Based on the atomic wavelength regions of~\citet{Smith2013} and ~\citet{Afsar2018}, 
we visually inspect several prominent atomic lines and compare them with synthetic spectra of which
chemical abundances are adjusted to match observed spectra. 
In order to avoid spurious results, the atomic lines that are very weak or significantly blended by other
lines are rejected.
The best matched spectrum with a minimum $\chi^2$-value between the synthetic and the observed spectra is
determined. The average of individual measurements and the standard deviation are decided as the
final chemical abundances and errors.
The estimated metallicities and abundances of Mg, Al, and Si are summarized in Table~\ref{tab:table3} with
respect to the solar abundances from~\citet{Asplund2009}. 

Calculated abundances are significantly affected by the uncertainty of atmospheric parameters. Therefore,
we comapre our atmospheric parameters with previous results of others, and quantify the abundance variation due to the parameter changes. 
Following the standard deviation of temperature relation of~\citet{Gonz2009},
the typical uncertainty in $T_{\mathrm{eff}}$ ($\Delta T$) is about $\sim100$ K, which leads average uncertainties in log $g$ and $v_{t}$ of
$\pm0.05$ dex and $0.02$ km s$^{-1}$, respectively.

We first compare our atmospheric parameters with those of~\citet{Boberg2015,Boberg2016} for the cross-matched stars 
in M53 and NGC 5053. The $T_{\mathrm{eff}}$ of~\citet{Boberg2015,Boberg2016} was calculated from the Alonso relation~\citep{Alonso1999}.
We find that the difference in $T_{\mathrm{eff}}$ with ~\citet{Boberg2015,Boberg2016} is about $\sim70$ K hotter with a standard deviation of  50 K, and only 
a few stars are hotter by $\sim150$ K.
The average differences in log $g$ and $v_t$ are small;  $0.08$ $(\sigma=0.12)$ dex and $0.04$ $(\sigma=0.17$) km s$^{-1}$, respectively. 
We note that the uncertainties of atmospheric parameters of~\citet{Boberg2015,Boberg2016} are 100 K in $T_{\mathrm{eff}}$,
0.2 dex in log $g$, and 0.25 km s$^{-1}$ in $v_t$, which indicates that the adopted atmospheric parameters
in this study are consistent with previous parameters within the error range.

We then investigate the sensitivity of abundances due to the variations in adopted atmospheric parameters. 
Abundances are re-estimated with new synthetic models, which are calculated by varying atmospheric parameters (temperature, gravity, and
microturbulence) one by one. The parameter changes are the uncertainties in the parameters 
(i.e., $\Delta T=100$ K, $\Delta \mathrm{log}$ $g=0.05$, and $\Delta v_{t}=0.02$ km s$^{-1}$).
Table~\ref{tab:table4} shows the mean sensitivity of abundances according to the atmospheric parameters changes.
The variation of atmospheric parameters results in a total uncertainty of about 0.1 dex in abundances; 
the effective temperature uncertainties are the main contribution of the abundance uncertainties.

\begin{table}
\begin{center}
\caption{Sensitivity of abundances due to variation of atmospheric parameters }
\label{tab:table4}
\begin{tabular}{cccc} 
\hline \hline  
Element & $\Delta T_{\mathrm{eff}} (\pm100 K) $ & $\Delta$ log $g$ ($\pm$0.05) & $\Delta v_t (\pm$0.02) \\
\hline 
Fe & $\pm0.07$ & $\mp0.07$ & $\pm0.03$ \\
Mg & $\pm0.08$  & $\mp0.01$ & $\pm0.01$ \\
Al & $\pm0.08$ & $\mp0.01$ & $\pm0.02$ \\
Si & $\pm0.09$ & $\pm0.05$ & $\pm0.03$ \\
\hline
\end{tabular}
\end{center}
\end{table}

\subsection{Abundance results}
From the manual abundance analysis, metallicities and elemental abundances of 73 cluster member candidates 
were investigated. 
Excluding stars with poor spectral quality, reliable estimates of metallicities and abundances of 65 stars (44 stars for
M53 and 21 stars for NGC 5053) were obtained, including six out of seven extra-tidal stars.
The average metallicities of M53 and NGC 5053 are [Fe/H]=-2.00$\pm$0.10 
and [Fe/H]=-2.17$\pm$0.07, which agrees with the metallicities that were previously reported by~\citet{Harris1996}. 
Here, we note that the 21 sample stars of NGC 5053 is the largest sample for which chemical abundances have been investigated in near-infrared high-resolution spectroscopy for this metal-poor cluster.

The left panel of Figure~\ref{fig:abun} shows the distribution of Al abundances as functions of Mg for the cluster stars. It is apparent that there are clear Mg-Al anticorrelations for M53 and NGC 5053. 
Al abundances of M53 show a large spread of about 1.0 dex, while those of NGC 5053 are about 0.8 dex.
We also found a clear gap in the Al distribution, which indicates that the populations of M53 and NGC 5053 are separated into two distinct 
abundance groups (i.e., Al-depleted first generation and Al-enhanced second generation). 
The extremely Mg-depleted stars ([Mg/Fe]$<$0), which are commonly detected in the most metal-poor globular clusters, such as M15 and M92~\citep{Masseron2019}, are not found in our clusters. 
Instead, the distribution of Mg abundance shows no strong variation for each cluster.
Mg-Al anticorrelation in globular clusters (including M53) has already been reported in many studies~\citep[e.g.,][]{Carretta2009,Meszaros2015,Masseron2019}.
Thus, we directly compared our results of M53 with those of~\citet{Masseron2019}, and found similar Mg-Al anticorrelation with almost the same Al abundance spread range.
However, the prominent Mg-Al anticorrelation of NGC 5053 studied in this study has never been investigated. 
A light symptom of Na-O and Mg-Al anticorrelation was reported~\citep{Boberg2015,Tang2018} for the
very limited samples of NGC 5053.
We found more first-generation stars of NGC 5053 near and below the upper limit of Al abundance by~\citet{Tang2018}, which
enabled us to find a clear Al variation in NGC 5053.

\begin{figure*}
\plottwo{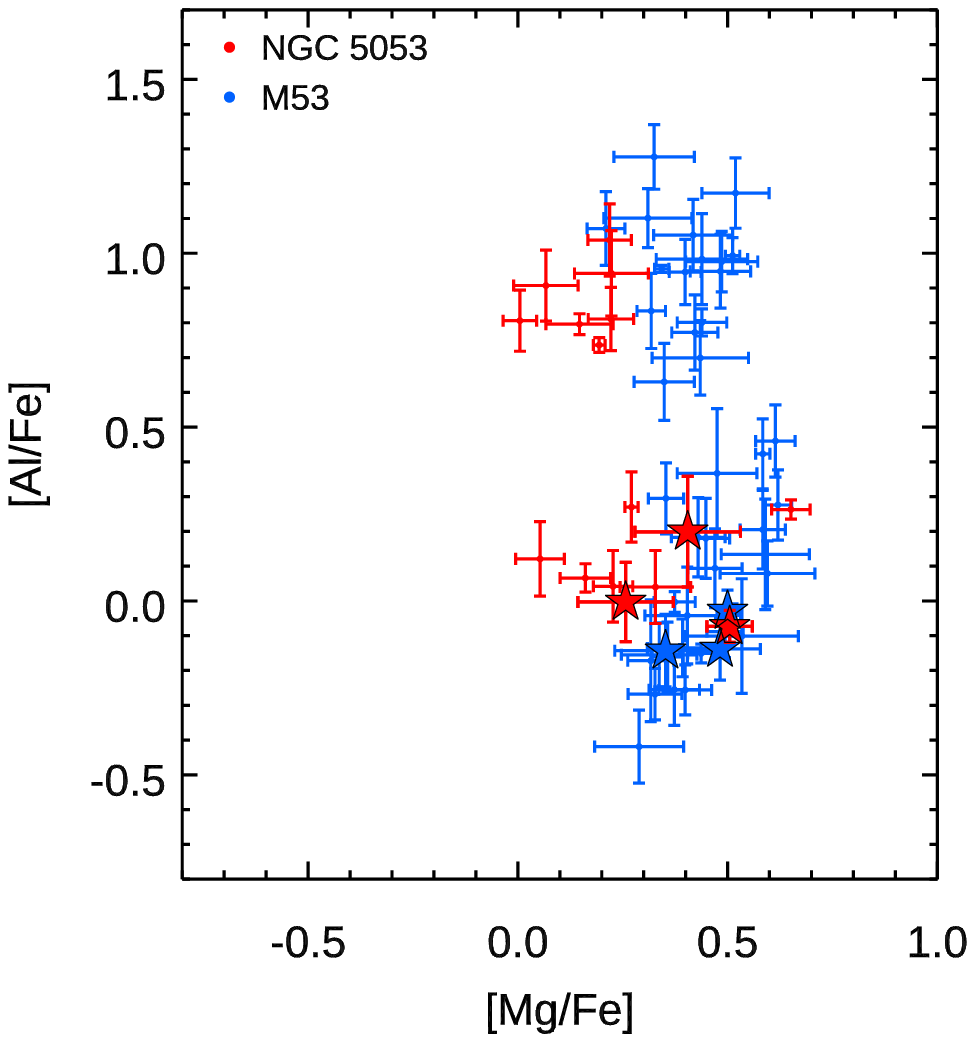}{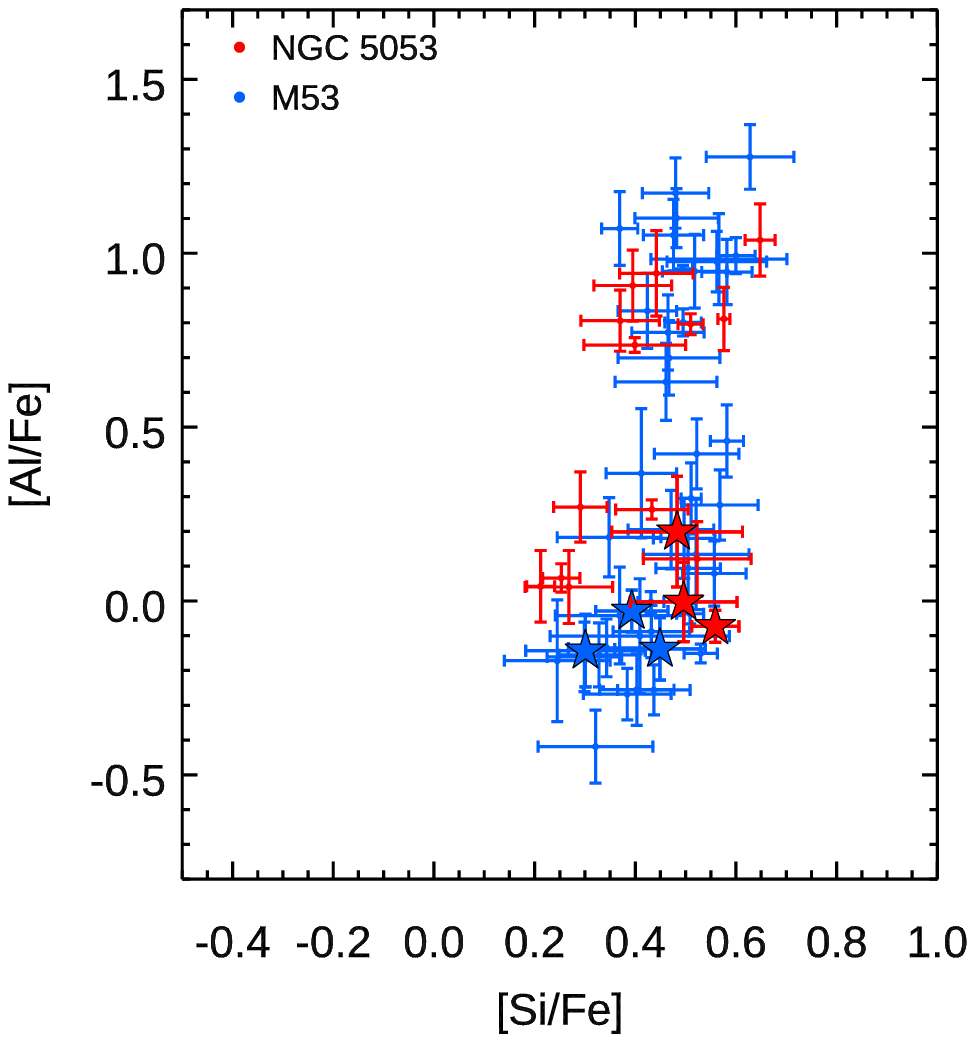}
\caption{Al abundances as function of Mg (left panel) and Si (right panel) abundances for cluster member stars. The blue and
red points are the stars of M53 and NGC 5053, respectively. The six extra-tidal stars are described by five point star marks. There is a 
clear Mg-Al anticorrelation for the stars of both clusters. The Al-Si plane shows the light Al-Si correlation.
}
\label{fig:abun}
\end{figure*}

The Al-Si distribution of the two clusters is plotted in the right panel of Figure~\ref{fig:abun}.
Our sample size of NGC 5053 is large enough and clearly confirms the Si spread in this cluster that was previously reported by~\citet{Tang2018}.
We found a light variation ($\sim0.3$ dex) in Si abundance for NGC 5053, which caused a light Al-Si correlation, 
while the Si abundance seemed to have a constant value for M53.
The Al-Si correlation was the result of $^{28}\mathrm{Si}$ leakage from the Mg-Al chain~\citep{Yong2005,Carretta2009,Meszaros2015}.
This nuclear reaction requires a very high temperature of $\sim80$ MK~\citep{Prant2017}, for which low-metallicity clusters or
massive clusters are preferable. The low-metallicity clusters M15 and M92 show an apparent Al-Si correlation with a significant
Si variation~\citep{Meszaros2015,Masseron2019}. NGC 5053 is also one of the most metal-poor clusters in the Milky Way with
a metallicity comparable to those of M15 and M92, even though its cluster mass is less massive. 
Therefore, a similar Si distribution of NGC 5053 to that of M15 and M92 is not surprising.
It is interesting that the less metal-poor but more massive cluster M53 does not show Si variation. 
In this respect, the different extents of Mg, Si, and Al variations in M53 and NGC 5053 indicates that 
metallicity is not the only factor that regulates the correlation between the light elements.

\section{Discussion and conclusion}
From the manual chemical analysis, Mg, Al, and Si abundances of six extra-tidal stars were reliably estimated.
We found that the identified extra-tidal stars had similar chemical properties with cluster stars (see Figure~\ref{fig:abun}), and 
they were all located in the low Al abundance region (i.e., first-generation of the clusters).
The Mg-rich/Al-poor feature of extra-tidal stars indicates that they could be the field stars that originated from first-generation populations of the clusters.
In globular clusters, first-generation stars are less centrally concentrated than second-generation stars~\citep{Lardo2011},
making them more vulnerable to tidal interaction. Thus, it is not surprising that most extra-tidal stars are first-generation stars.

The origin of multiple populations in globular clusters is still being debated, and no single model can successfully explain all the observational
results. Still, a common prediction among different models~\citep[e.g.,][]{Decressin2007,Dercole2008,Ventura2008,Bastian2013,Bastian2018} is that 
globular clusters were initially much more massive than currently observed, and a large fraction of stars have been lost 
since their formation. Some models suggest that globular clusters were 25 times more massive than present and lost as much as 90 $\%$ of their stars. 
Other studies~\citep{Kruijssen2015,Baumgardt2017,Baumsoll2017,Baumgardt2019} suggest that initial globular clusters are 4$\mathrm{-}5$ times larger and average star loss is about $75\mathrm{-}80\mathrm{\%}$. 
We note that recent studies~\citep{Vesperini2010,Larsen2012,Milone2017,Schi2017b} have indicated that heavy mass loss of up to 90 $\%$ is unrealistic and presents several problems~\citep{Bastian2015}, and that about 50 $\%$ mass loss seems more acceptable.
In this case, at least about 10 $\%$ of the Galactic halo stars could have originated
from globular clusters~\citep{Martell2016,Koch2019}.
Recently, ~\citet{Hanke2020} investigated the chemodynamical association of the halo stars with globular clusters, and suggested that the fraction of
first-generation cluster stars among all stars escaped from clusters into the halo is about 50$\%$ in the vicinity of the clusters and 80$\%$ in the
distant halo field.
Therefore, we propose that 
the extra-tidal stars that we found in this work were first-generation stars in M53 and NGC 5053 that became unbound from their parental clusters.

In this study, we found 73 cluster member candidates of two globular clusters, M53 and NGC 5053, using the t-SNE algorithm, radial velocity, and
a manual chemical abundance analysis. Out of those, seven stars were beyond the tidal radii of two clusters and were thus likely to be extra-tidal stars associated
with either M53 or NGC 5053.
The extra-tidal stars in the vicinity of the clusters appeared to share the proper motion direction with those of the clusters.
Furthermore, small clumps in stellar density contour of~\citet{Chun2010} and these stars were
well-aligned along the trailing and leading direction of cluster proper motion. 
The morphology of distortion in marginal stellar density contour seems to approach these stars.
A chemical abundance analysis for extra-tidal stars showed that these stars could be first-generation stars stripped from the two clusters
by tidal disruption or tidal interaction between the clusters.

It is notable that one extra-tidal star of NGC 5053 is in the M53 side of the common envelope of~\citet{Chun2010}.
This may represent the stars that were tidally stripped and gravitated toward its neighbor, and it may serve as a direct evidence that 
two clusters experienced dynamical interaction, although the radial velocities are different by 105 km s$^{-1}$. A tidal link between the two clusters could indicate that these clusters did not originate from the Milky Way, but from dwarf galaxies, as the interaction between the clusters would have occurred more preferentially in dwarf galaxies~\citep{Van1996}.
~\citet{Mackey2004} noted that M53 is an accreted cluster.
~\citet{Vasiliev2019} used 6d phase space information of all the globular clusters in the Milky Way from GAIA data and found that M53 and NGC 5053 have very similar dynamical structures together with several more globular clusters, (i.e., similar total energy, orbit and z-component angular momentum), which infers the possible accretion origin thereof from the dwarf galaxy. The Sgr dwarf galaxy was not their progenitor, as it shows a significantly different orbit from those of clusters~\citep{Sohn2018,Tang2018}. More recently, ~\citet{Massari2019} investigated the origin of globular clusters and tried to link the known merging or accretion events in the Milky Way using the same 6d phase space information;
 they further suggested that M53 and NGC 5053 belong to the Helmi streams~\citep{Helmi1999}.

In this respect, finding more extra-tidal stars that could reflect tidal interaction between the clusters, especially in the tidal
bridge or common envelope of~\citet{Chun2010}, is important to better understand the dynamical evolution of these clusters.
Four extra-tidal RR Lyrae stars of M53~\citep{Kundu2019} located inside the tidal radius of NGC 5053 could be of interest,
but we are certain that these stars are the member stars of NGC 5053, as reported by~\citet{Ngeow2020}. ~\citet{Ngeow2020} searched for additional
RR Lyrae stars in the vicinity of the two clusters, and concluded that there are no extra-tidal RR Lyrae stars associated with either M53 or NGC 5053.
However, we still expect that the near-field surrounding these clusters has the potential to find more extra-tidal stars.
The present RR Lyrae stars in the clusters had been more massive than RGB and main-sequence stars, while the low mass stars are more easily affected by a tidal stripping event or tidal interaction. The APOGEE stars explored in this study are also very bright RGB stars in the two clusters. 
Therefore, the detection of our extra-tidal RGB stars with a bright magnitude increases the possibility of finding more extra-tidal stars with a fainter magnitude.

In summary, we note that additional photometry and spectroscopy studies for these clusters and the surrounding stars are required to find
more definitive evidence of the tidal disruption and the tidal interaction between them. 
Since RGBs only make up a small percentage of the stellar populations in globular clusters, it is necessary to search for any tidal substructures of 
more numerous populations, such as main-sequence stars.
We note that the most populations that comprise the stellar substructures around M53 and NGC 5053 detected by~\citet{Chun2010} are, indeed, main-sequence stars. Homogeneous deep and wide photometry data could provide a finer morphology of stellar substructures to infer the tidal disruption and a possible link between the clusters. The follow-up spectroscopic data of SDSS-V~\citep{Koll2019}, 4MOST~\citep{Christ2019,Helmi2019}, and MOONS~\citep{Cira2012} for fainter stars than APOGEE samples could also provide more reliable chemical associations for extra-tidal stars in the substructures of the clusters. Numerical simulations of binary star clusters are also essential to understand the kinematics of tidally stripped stars by neighboring clusters.

\acknowledgments
This work was supported by Korean GMT project operated by Korea Astronomy and Space Science Institute (KASI).

DL acknowledges support from the Deutsche Forschungsgemeinschaft (DFG, German Research Foundation) -- Project-ID 138713538 -- SFB 881 (``The Milky Way System'', subproject A03).

Funding for the Sloan Digital Sky Survey IV has been provided by the Alfred P. Sloan Foundation, the U.S. Department of Energy Office of Science, and the Participating Institutions. SDSS-IV acknowledges
support and resources from the Center for High-Performance Computing at
the University of Utah. The SDSS web site is www.sdss.org.

SDSS-IV is managed by the Astrophysical Research Consortium for the 
Participating Institutions of the SDSS Collaboration including the 
Brazilian Participation Group, the Carnegie Institution for Science, 
Carnegie Mellon University, the Chilean Participation Group, the French Participation Group, Harvard-Smithsonian Center for Astrophysics, 
Instituto de Astrof\'isica de Canarias, The Johns Hopkins University, Kavli Institute for the Physics and Mathematics of the Universe (IPMU) / 
University of Tokyo, the Korean Participation Group, Lawrence Berkeley National Laboratory, 
Leibniz Institut f\"ur Astrophysik Potsdam (AIP),  
Max-Planck-Institut f\"ur Astronomie (MPIA Heidelberg), 
Max-Planck-Institut f\"ur Astrophysik (MPA Garching), 
Max-Planck-Institut f\"ur Extraterrestrische Physik (MPE), 
National Astronomical Observatories of China, New Mexico State University, 
New York University, University of Notre Dame, 
Observat\'ario Nacional / MCTI, The Ohio State University, 
Pennsylvania State University, Shanghai Astronomical Observatory, 
United Kingdom Participation Group,
Universidad Nacional Aut\'onoma de M\'exico, University of Arizona, 
University of Colorado Boulder, University of Oxford, University of Portsmouth, 
University of Utah, University of Virginia, University of Washington, University of Wisconsin, 
Vanderbilt University, and Yale University.

\bibliographystyle{aasjournal}
\bibliography{reference}
\end{document}